\documentclass[aps,prl,reprint,groupedaddress,twocolumn]{revtex4-1}

\usepackage{graphicx}
\usepackage{dcolumn}
\usepackage{bm}
\usepackage{amsmath,amssymb,subfigure}
\usepackage{color}
\usepackage[normalem]{ulem}
\usepackage{epstopdf}

\newcommand{\eq}{Eqn.~}
\newcommand{\etal}{\textit{et al.}}
\newcommand{\bnd}[1]{\lfloor #1 \rceil}
\newcommand{\floor}[1]{\lfloor #1 \rfloor}
\newcommand{\ceil}[1]{\lceil #1 \rceil}
\newcommand{\gammasl}{\bnd{\gamma_{sl_0}}}
\newcommand{\gslfloor}{\floor{\gamma_{sl_0}}}
\newcommand{\gslceil}{\ceil{\gamma_{sl_0}}}

\begin{document}


\title{Observation of contact angle hysteresis due to inhomogeneous electric fields}


\author{Wei Wang}
\author{Qi Wang}
\author{Jia Zhou}
\email{jia.zhou@fudan.edu.cn}
\author{Antoine Riaud}
\email{antoine\_riaud@fudan.edu.cn}
\affiliation{State Key Laboratory of ASIC and System, School of Microelectronics, Fudan University, Shanghai 200433, China}


\date{\today}

\begin{abstract}
	Static contact angle hysteresis (CAH) is widely attributed to surface roughness and chemical contamination. In the latter case, chemical defects create free-energy barriers that prevent the contact line motion. Electrowetting studies have demonstrated the similar ability of electric fields to alter the surface free-energy landscape. Yet, the increase of apparent static CAH by electric fields remains unseen. Here, we report the observation and theoretical analysis of electrowetting hysteresis. This phenomenon enables the continuous and dynamic control of CAH, not only for fundamental studies but also to manufacture sticky-on-demand surfaces for sample collection.
\end{abstract}


\maketitle



On most surfaces, the liquid-solid contact angle $\theta$ can adopt a range of values before the contact line starts moving \cite{Gennes1985,Eral2013,Erbil2014}. This static contact angle hysteresis (CAH, the difference between the advancing and receding contact angles) is mainly attributed to surface roughness \cite{Johnson1964,cox1983spreading} and chemical heterogeneities \cite{Johnson1964b,Priest2007}. Despite its major implications \cite{Eral2013}, CAH remains highly challenging to study experimentally because it is a macroscopic phenomenon excessively sensitive to nanometer-scale defects. Hence, experimental surfaces need to be prepared with the greatest care and quantitative well-controlled experiments are scarce \cite{chatain1995experimental,Priest2007}.  In most cases, patches of a selected contaminant are carefully deposited on an ultra-clean polished surface. There is a widespread consensus that the contaminant creates free-energy barriers that prevent the contact line motion \cite{Good1952,Gennes1985}. This modification of the free-energy landscape is well described by the Gibbs isotherm $\frac{\partial \gamma_0}{\partial \mu_i} = -\Gamma_i$, with $\Gamma_i$ the surface coverage of a contaminant $i$, $\mu_i$ its chemical potential, and $\gamma_0$ the local surface tension. Hence, once the composition of the surface is set, the surface energy landscape is not allowed to vary anymore, which makes CAH experiments extremely work-intensive.

Over the last two decades, electrowetting has emerged as a convenient way to control the apparent contact angle of liquid droplets by adjusting the effective liquid-solid interfacial tension using electric fields \cite{sun2019surface,mugele2005electrowetting,Fair2007,Nelson2012,Heikenfeld2009,Hayes2003,Li2012,Hao2014,Heikenfeld2005,Murade2011}:

While electrowetting was originally explained by Lippmann on thermodynamic grounds \cite{Lippmann1875}, Jones \cite{jones2002relationship} and Buehrle \etal  \cite{Buehrle2003} have clarified that the liquid-solid interfacial tension and the contact angle are not affected by the electric fields \cite{Bonn2009}, but that the interface profile is gradually evolving from the intrinsic Young-Dupr\'e contact angle to the apparent Young-Lippmann contact angle \cite{mugele2007equilibrium}. This transition length depends on the electrowetting setup but is generally negligible compared to the droplet size as it most often occurs within 1 $\mathrm{\mu m}$ from the contact line. 

Outside this transition region, the thermodynamic identity derived by Lippmann is recovered by posing the effective interfacial tension $\gamma = \frac{\partial F}{\partial A}$ as the generalized free energy per unit solid-liquid surface area $A$ \cite{israelachvili2011intermolecular}. The generalized free-energy $F = E - Q\Phi - \sum_{i} \mu_i n_i$ offsets the Helmholtz free-energy $E$ so that it is minimized at thermodynamic equilibrium under fixed electric potential $\Phi$ and chemical potential, regardless of the amount $n_i$ of species $i$ and of the electric charge $Q$. Using Maxwell identities on $F$ yields the celebrated Lippmann equation (additional discussion and derivation available in SI):
\begin{equation}
\frac{\partial \gamma}{\partial \Phi} = -\sigma,
\label{eq: Lippmann0}
\end{equation}
with $\sigma$ the areal surface charge. The striking analogy between \eq\ref{eq: Lippmann0} and the Gibbs isotherm suggests that surface charges and electric fields may also influence the CAH on a scale larger than the electrowetting transition region. Such electrical control of CAH may open exciting strategies for liquid collection and release by reversibly switching the surface between sticky and non-sticky states, respectively. More broadly, it might enable quantitative CAH studies that (i) do not rely on defects and (ii) investigate a continuous range of advancing and receding conditions. To the best of knowledge, this electrically-induced CAH has never been observed. Moreover, most studies in electrowetting phenomena report either no effect \cite{nelson2011dynamic,mchale2011,bhushan2011role} or even a reduction \cite{li2008make,nita2017electrostatic} of CAH due to the application of an electric voltage. 

These confusing observations are mainly due to three reasons. First, instead of the static CAH \cite{Eral2013,Gennes1985,Erbil2014} observed at mechanical equilibrium, most electrowetting studies are concerned with moving droplets and therefore focus on the more complex dynamic CAH, which involves viscous and inertial forces that may affect the hysteresis \cite{nelson2011dynamic,li2008make}. Second, by analogy with chemical heterogeneities and surface roughness, the electrical control of CAH should require inhomogeneous electric fields, which were used only in a handful of reports \cite{Yi2006,mchale2011}. We note that homogeneous electric fields can increase the droplet friction by expanding the three-phase contact perimeter \cite{wang2009liquid} or may also alter the CAH by triggering a wetting state transition \cite{manukyan2011electrical}, but the CAH in each state is then controlled by the surface roughness and not by the electric field. Third, the definition of hysteresis differs between electrowetting and the usual wetting phenomena due to chemical heterogeneities and surface roughness. Indeed, the CAH is obtained by recording the onset of contact line motion, which can be achieved in two ways in electrowetting studies: (i) by hydrostatic stress (such as inclining the plane or pumping the liquid) or (ii) by electrocapillary stress \cite{Eral2013,Yi2006,mchale2011}. Since the latter has no direct analog in the chemical and roughness-induced hysteresis, it does not allow to conclude on the ability of electric fields to modify the CAH. Hence, the conditions to observe the variation of CAH due to an electric field have not yet been fulfilled.

\begin{figure*}[htp]
	\includegraphics[width=17.2cm]{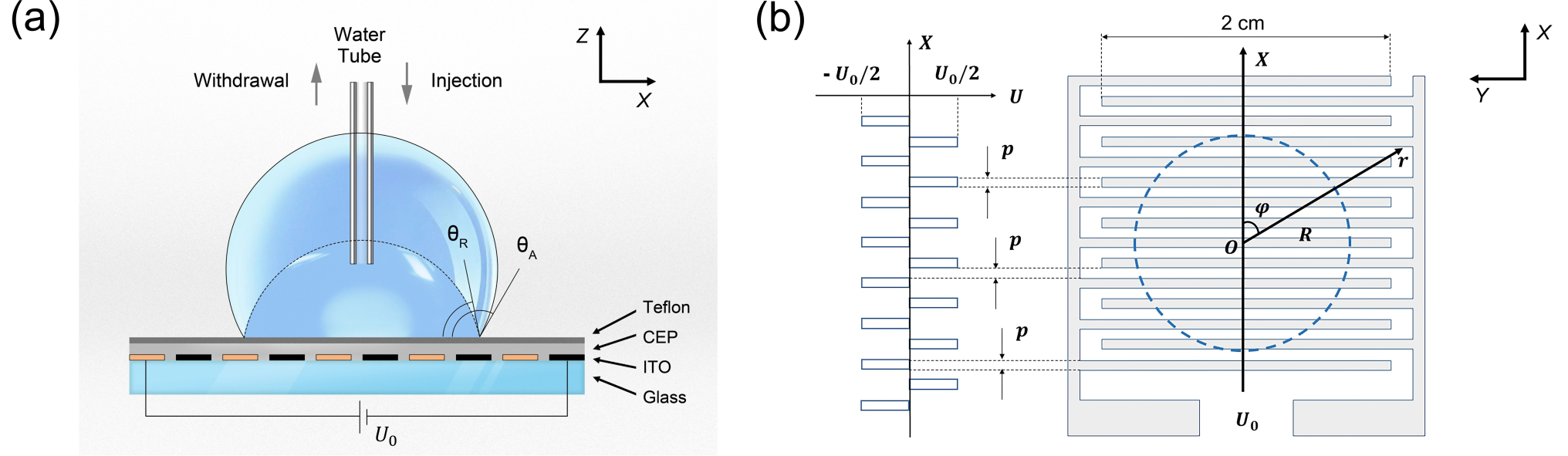}
	\caption{Experimental setup. (a) Side view. The liquid is replenished or extracted via a steel tubing. The indium tin oxide (ITO) interdigitated electrodes (IDE) are isolated from the liquid by a dielectric layer of cyanoethyl pullulan (CEP) and a hydrophobic layer of Teflon, and connected to a DC supply $U_0$. Side-view images were captured and analyzed using a goniometer. (b) Left: Voltage across the solid-liquid interface in $x$ direction. Due to the large aspect ratio of the IDEs, the electric potential is well approximated by a square-wave function \cite{Morgen2001,Gunda2010} with $p$ the finger width and spacing. Right: Top view. The contact line radius of the droplet is denoted by $R$. In these polar coordinates, the origin $O$ is located at the center of the circular contact area (dashed circle), with $\varphi$ and $r$ the angular and radial coordinates, respectively.}
	\label{Setup}
\end{figure*}

Here, we demonstrate the direct analogy of static CAH by applying an inhomogeneous electric field while pumping liquid in the droplet. A thermodynamic model is derived to describe the evolution of advancing and receding apparent contact angles. Similarly to earlier works on CAH \cite{Joanny1984,Good1952,Johnson1964b}, our model overlooks the finest details of the contact line structure such as the precursor wetting film \cite{yuan2010precursor}, Van der Waals force \cite{huh1977effects} and Lorentz force acting at the interface \cite{Buehrle2003}. Despite these simplifications, it captures well the key features of the experimental observations, including a quadratic dependence on the actuation voltage and an unexpected influence of the droplet contact line radius. 


The experimental setup \cite{wang2020demand}, shown in Fig.~\ref{Setup}, is similar to the dielectrowetting chip used previously by McHale \etal ~\cite{mchale2011}. A periodic array of indium-tin-oxide (ITO) interdigitated electrodes (IDEs, 50 $\mathrm{\mu m}$ finger width and spacing, 130 nm thick) ensures the generation of an inhomogeneous electric field, which is fundamental to observe a CAH. This field is reminscent of the stripes  used in several studies of CAH (see \cite{choi2009modified} and references therein). A dielectric cyanoethyl pullulan layer (408 nm thick) insulates the liquid from the electrodes \cite{chen2014study} so that the whole droplet behaves like an ideal conductor (dielectric relaxation time $\tau_e=\epsilon_l/\sigma_l\simeq1.3\times10^{-4}$ s, with $\epsilon_l\simeq80\epsilon_0$ and $\epsilon_0=8.85\times10^{-12}$ F/m the dielectric permittivity of water and vacuum respectively, and $\sigma_l \leq 5.6\times10^{-6} $ S/m the conductivity of deionized water). Since electrowetting can only reduce the contact angle, a thin hydrophobic polytetrafluoroethylene (PTFE) layer (60 nm thick) is coated on the dielectric layer to maximize the operating range. A steel tubing is used to replenish or extract liquid during CAH measurements. The tubing is not electrically connected.

For a given voltage, a 40 $\mathrm{\mu L}$ droplet is deposited on the chip and spreads by electrowetting until it reaches a stable shape. Then, the droplet volume is slowly varied by pumping liquid via the tubing. The relatively low flow rate  $Q_l=0.1$ $\mathrm{\mu L/s}$ ensures (i) that the droplet behaves as an ideal conductor (the characteristic pumping time $\tau_p = V_l/Q_l\simeq 400$ s $ \gg \tau_e $ with $V_l$ the droplet volume) and (ii) that the droplet shape is always near mechanical equilibrium ($\tau_{cap} = \sqrt{\frac{\rho_lV_l}{\sigma}}\simeq$ 23 ms $\ll\tau_p$ with $\tau_{cap}$ the mechanical relaxation time, $\rho_l\simeq1000$ kg/m$^3$ and $\sigma\simeq72.8\times10^{-3}$ the density and surface tension of water in air) and dynamic CAH can be neglected. Due to the CAH, the contact line remains trapped during the pumping cycles until the deformation overcomes the static CAH. Hence, advancing and receding contact line radii are the same for a constant voltage. Side view images of the droplet are captured to measure contact angles with a goniometer (DSA30, KR$\mathrm{\ddot{U}}$SS, Germany). After each acquisition, the surface is cleaned, and the experiment is repeated with a different voltage.

\begin{figure}[tbp]
	\includegraphics[width=8.6cm]{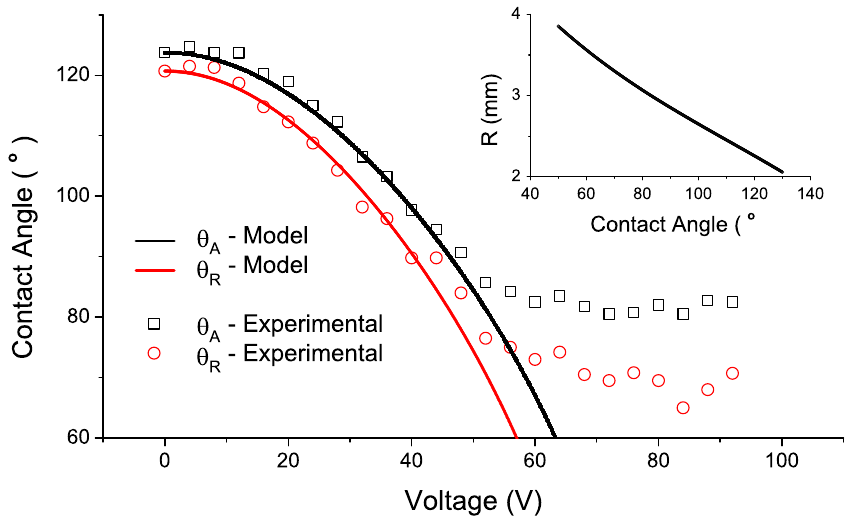}
	\caption{Effect of the electric voltage on apparent contact angles. The experimental observations agree well with the model (\eq\ref{eq: costhetaR} and \eq\ref{eq: costhetaA}) in the low-saturation region ($|U_0|< 50$ $\mathrm{V_{DC}}$). Each experimental dot was averaged over 5 tests and error bars are not presented for the readability. Inset: Contact line radius $R$ versus contact angle for 40 $\mathrm{\mu L}$ droplets used in our experiments. It is determined by numerically solving the differential Young-Laplace equation \cite{Lubarda2011}.}
	\label{Experiments-Theory}
\end{figure}

The measured apparent contact angles for $U_0$ ranging from -100 to +100 $\mathrm{V_{DC}}$ are reported in Fig.~\ref{Experiments-Theory}. In agreement with the celebrated Lippmann-Young equation and other electrowetting experiments \cite{mugele2005electrowetting,Nelson2012,Mugele2002}, the contact angle decreases with the voltage until $|U_0|$ reaches approximately $50$ $\mathrm{V_{DC}}$, above which electrowetting saturation occurs. Given the controversy surrounding the origins of electrowetting saturation \cite{liu2012uncovering,klarman2011model,chevalliot2012experimental,papathanasiou2008illuminating,peykov2000electrowetting}, we prefer to restrict this study to the low-voltage region ($|U_0|< 50$ $\mathrm{V_{DC}}$). 

\begin{figure}[tbp]
	\includegraphics[width=8.6cm]{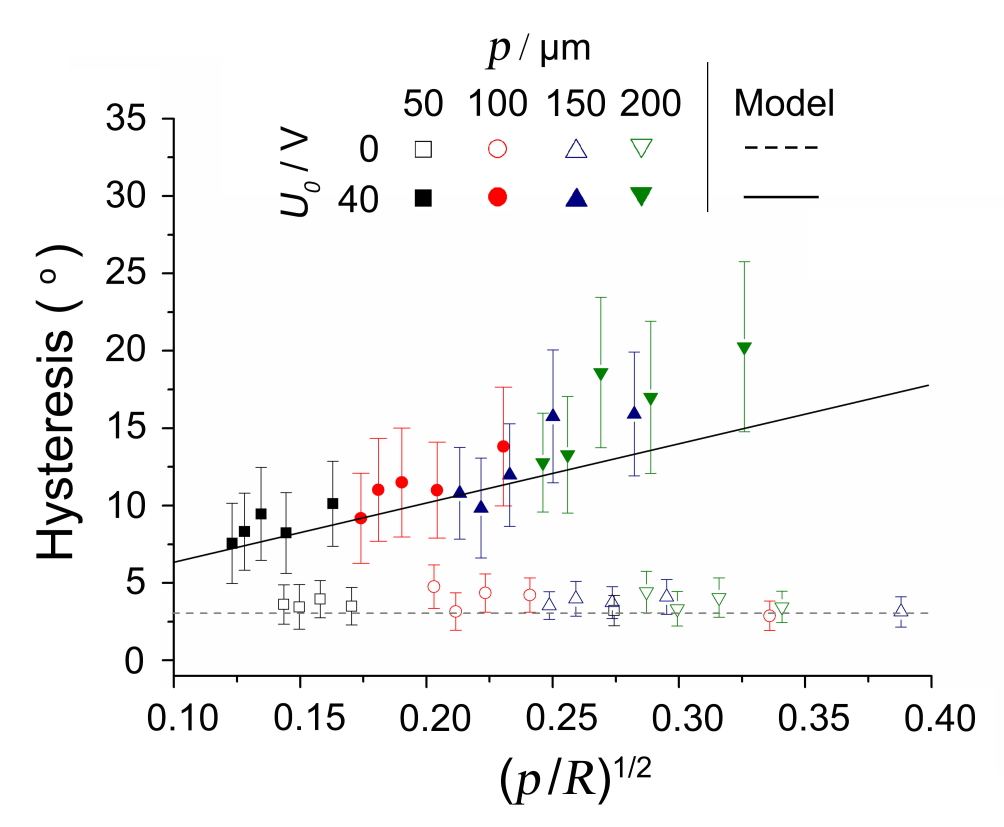}
	\caption{Contact angle hysteresis measured at $0$ and $40$ V with electrode pitches ranging from $50$ to $200$ $\mu$m and droplet volumes ranging from $10$ to $50$ $\mu$ L. Droplets of larger volumes show smaller hysteresis. The value of $R$ was estimated by solving the Laplace-Young equation. Each experimental dot was averaged over 5 tests. The model was obtained by taking $C = 0.350$ mF/m$^2$, $\gamma_{lg}=71.97$ mN/m, $\theta_{A0}=123.8^o$, and $\theta_{R0}=120.7^o$.}
	\label{Hysteresis}
\end{figure}

The static CAH deduced from Fig.~\ref{Experiments-Theory} is shown in Fig.~\ref{Hysteresis}. Unlike other studies \cite{nelson2011dynamic,mchale2011,bhushan2011role,li2008make,nita2017electrostatic}, we observe an increase of CAH (from 3.1$\mathrm{^o}$ to 7.5$\mathrm{^o}$) as $|U_0|$ increases from $0$ to $50$ $\mathrm{V_{DC}}$. At this point, the CAH features a marked dip before growing far into the saturation regime. The exact reasons for this dip are not yet known but its collocation with the electrowetting saturation suggests that both phenomena might be linked. 

In the following, we model the static CAH below saturation regime as a function of the actuation voltage within the framework developed by Johnson and Dettre \cite{Johnson1964b}. The apparent contact angle is obtained by finding the apparent contact line radius $R$ that minimizes the generalized free energy $F = F_{sl} + F_{lg} + F_{sg}$ of the system, with $F_{sl}$, $F_{lg}$, and $F_{sg}$ the generalized free energies of the solid-liquid, liquid-gas, and solid-gas interfaces, respectively. The elementary displacement $\partial R$ is taken much smaller than the IDE width \cite{choi2009modified} but much larger than the transition region (approximately the combined thickness of the hydrophobic and dielectric layers, that is 0.5 $\mu$m) so that variations of generalized free energy at the contact line are negligible compared to the variations of solid-liquid free energy. Assuming that hysteresis is weak enough so that the droplet remains circular at all times, which was verified for two orthogonal directions (data available in SI), geometrical construction yields $\partial F_{lg}/\partial R = 2\pi R\gamma_{lg}\cos\theta$ and $\partial F_{sg}/\partial R=-2\pi R\gamma_{sg}$, with $\gamma_{lg}$ and $\gamma_{sg}$ the liquid-gas and solid-gas interfacial tensions, respectively \cite{Roura2004}. The derivation of $\partial F_{sl}/\partial R$, which differs from the classical Young-Dupr\'e equation \cite{Roura2004}, is our main concern.

By definition, $F_{sl}=\epsilon_{CL}+\mathcal{C}+\int_{0}^{R}\int_{-\pi}^{\pi}\gamma_{sl} r \mathrm{d}r \mathrm{d}\varphi$ with $\epsilon_{CL}$ the free energy in the transition region, $\mathcal{C}$ a constant, $\gamma_{sl}$ the solid-liquid effective interfacial tension, and $\varphi$ and $r$ the angular and radial coordinates, respectively (see Fig.~\ref{Setup}). Because the transition region is much smaller than the elementary volume that we are studying, $\epsilon_{CL}$ is much smaller than $F_{sl}$. Elementary calculus yields:
\begin{equation}
\frac{\partial F_{sl}}{\partial R}=R\int\limits_{-\pi}^{\pi}\gamma_{sl} \mathrm{d}\varphi.
\label{eq: dEsl0}
\end{equation}
We evaluate  $\gamma_{sl}$ by assuming that the electric field varies continuously at the microscale, and can thus be locally integrated from \eq\ref{eq: Lippmann0}:
\begin{equation}
\gamma_{sl} = \gammasl-\frac{C}{2}U^2,
\label{eq: Lippmann}
\end{equation}
with $U$ and $C$ the effective voltage across the solid-liquid interface and the areal capacitance between the electrodes and the liquid (this parameter is determined experimentally in SI). $\gammasl$ is the intrinsic interfacial tension, that is $\gamma_{sl}$ in the absence of electric field. The $\bnd{}$ notation indicates that, due to chemical impurities \cite{Gennes1985}, $\gammasl$ is susceptible to vary between a lower and upper bound given by $\gslfloor$ and $\gslceil$ respectively. According to Young-Dupr\'e equation \cite{Eral2013}, the advancing contact angle $\theta_{A0}$ and receding contact angle $\theta_{R0}$ at zero-voltage satisfy $\cos\theta_{A0}=(\gamma_{sg}-\gslceil)/\gamma_{lg}$ and $\cos\theta_{R0}=(\gamma_{sg}-\gslfloor)/\gamma_{lg}$, respectively.

For a regular array of IDEs as shown in Fig.~\ref{Setup}(b), the effective voltage $U$ is half of the externally applied voltage $U_0$ (see SI for the full derivation):
\begin{equation}
U^2=\left(\frac{U_0}{2}\right)^2 \Pi(kx),
\label{eq: voltage}
\end{equation}
where $\Pi$ is the square-wave function denoting the distribution of the electric potential energy ($1$ above the electrodes and $0$ elsewhere, similar to the hysteresis-prone mesa-type landscape of Joanny and De Gennes \cite{Joanny1984}) and $k$ denotes the wavenumber $\pi/p$ with $p$ the finger width and also the spacing (see Fig.~\ref{Setup}) \cite{Morgen2001,Gunda2010}.  Integration of \eq\ref{eq: dEsl0} is simpler when $\Pi$ is expressed as a Fourier series:
\begin{equation}
\Pi=\frac{1}{2}+\sum\limits_{n=0}^{\infty}\frac{2(-1)^n}{(2n+1)\pi}\cos[(2n+1) k r\cos\varphi].
\label{eq: indicator}
\end{equation}

Substituting \eq\ref{eq: Lippmann}, \ref{eq: voltage}, and \ref{eq: indicator} in \eq\ref{eq: dEsl0} and using the identity $\int_{-\pi}^{\pi}\cos(\tau\cos\varphi)d\varphi=2\pi J_0(\tau)$ with $J_0$ the 0-order Bessel function, we get:
\begin{subequations}
	\begin{align}
	{\frac{\partial F_{sl}}{\partial R}}
	&=2\pi R\left\{\gammasl-\gamma_{L} - \gamma_H\right\},
	\label{eq: dEsl}\\
	\gamma_{L} &= \frac{CU_0^2}{16}, \label{eq: gammaL}\\ 
	\gamma_{H} &= \frac{CU_0^2}{8}\sum_{n=0}^{\infty}\frac{2(-1)^n}{(2n+1)\pi}J_0[(2n+1)kR]. \label{eq: gammaH}
	\end{align}
\end{subequations}
It contains three terms: the intrinsic interfacial tension $\gammasl$, the average decrease in effective interfacial tension $\gamma_L$ (see \cite{Yi2006} for IDEs), and an oscillating term $\gamma_{H}$ that results in an effective hysteresis effect but did not appear in previous studies. For uniform potential distributions ($k\rightarrow0$), \eq\ref{eq: dEsl}  reduces to the standard Lippmann-Young equation. 
In the current experiments, $R \gg p$,  so $J_0(\tau) \simeq \sqrt{\frac{2}{\pi\tau}}\cos (\tau-\frac{\pi}{4})$, which simplifies \eq\ref{eq: gammaH} to 
\begin{equation}
\gamma_{H}
\approx \frac{CU_0^2}{8\sqrt{\pi kR}}\sum_{n=0}^{\infty}\frac{2\sqrt{2}(-1)^n}{\pi(2n + 1)^{3/2}}
\cos \left[(2n+1)kR- \frac{\pi}{4}\right].
\label{eq: gammaH_approx}
\end{equation}
In the SI, we postulate that the upper and lower bounds of $\gamma_{H}$ read:
\begin{equation}
-B\frac{CU_0^2}{16}\sqrt{\frac{p}{R}} < \gamma_{H} < B\frac{CU_0^2}{16}\sqrt{\frac{p}{R}},
\label{eq: gamma_H_bnds}
\end{equation}
with $B = \frac{4-\sqrt{2}}{\pi^2}\zeta\left(\frac{3}{2}\right) \approx 0.684$ and $\zeta$ the Riemann zeta function. Interestingly, \eq\ref{eq: gamma_H_bnds} indicates the dependence of static CAH on $\sqrt{\frac{p}{R}}$. 

Finally, substituting \eq\ref{eq: gamma_H_bnds} in \eq\ref{eq: dEsl} and minimizing the generalized free energy $F$ yields: 
\begin{subequations}
	\begin{align}
	\cos \theta_A &< \cos \theta <\cos \theta_R 
	\label{eq: CAH}\\
	\cos \theta_A &= \cos \theta_{A0} + \frac{CU_0^2}{16\gamma_{lg}}-B\frac{CU_0^2}{16\gamma_{lg}}\sqrt{\frac{p}{R}} \label{eq: costhetaA}\\
	\cos \theta_R &= \cos \theta_{R0} + \frac{CU_0^2}{16\gamma_{lg}}+B\frac{CU_0^2}{16\gamma_{lg}}\sqrt{\frac{p}{R}} \label{eq: costhetaR}
	\end{align}
	\label{model}
\end{subequations}
The upper and lower bounds of the contact angle in \eq\ref{eq: CAH} correspond to the receding apparent contact angle $\theta_R$ and the advancing apparent contact angle $\theta_A$, respectively.

Subtracting \eq\ref{eq: costhetaA} and \eq\ref{eq: costhetaR}, and assuming that the CAH remains small yield:
\begin{subequations}
	\begin{align}
	\delta\cos \theta_R - \delta \cos\theta_A = B\frac{CU_0^2}{8\gamma_{lg}}\sqrt{\frac{p}{R}} \label{eq: deltacostheta} \\
	\theta_A-\theta_R = \frac{\gslceil - \gslfloor}{\gamma_{lg}\sin\theta_E}+\frac{BCU_0^2}{8\gamma_{lg}\sin\theta_E}\sqrt{\frac{p}{R}}
	\label{eq: ewod CAH}
	\end{align}
\end{subequations}
with $\delta\cos \theta_i = \cos \theta_i - \cos \theta_i(U=0)$ for $i=R$ or $A$, and $\theta_E=(\theta_{A}+\theta_{R})/2$. \eq\ref{eq: deltacostheta} provides an energetic viewpoint that singles out the electrical effect on the effective contact angle hysteresis, while \eq\ref{eq: ewod CAH} is a convenient expression for experimental purposes. The first term in \eq\ref{eq: ewod CAH} describes the hysteresis due to surface defects, while the second represents the electrowetting hysteresis. We note that $1/\sin(\theta_E)$ is a decreasing function of the actuation voltage on the studied interval ($0$ to $40$ V) and therefore cannot be responsible for the observed increase in CAH (the contribution of both terms is shown in SI).


The predictions of \eq\ref{model} are compared to experimental results in Fig.~\ref{Experiments-Theory}. The areal capacitance $C\simeq0.350$ mF/$m^2$ (see SI) and the advancing and receding contact angles in the absence of electric field $\theta_{A0}\simeq123.8^o$ and $\theta_{R0}\simeq120.7^o$ are determined by fitting the experimental data. The droplet contact line radius, required in \eq\ref{model}, is estimated by solving numerically the Young-Laplace equation (see SI). Similarly to previous electrowetting on dielectric (EWOD) studies  \cite{mugele2005electrowetting,Nelson2012,Mugele2002}, theory and experiments agree well within the non-saturated region. 

Next \eq\ref{eq: ewod CAH} is compared to experimental data in Fig.~\ref{Hysteresis}. For the sake of clarity, the comparison is limited to  $0$ and $40$ V for droplet volumes ranging from $10$ to $50$ $\mu$L and electrode pitches from $50$ to $200$ $\mu$m (comparison over a broader range of voltages is available in SI). These experiments were carried out immediately after manufacturing the substrates, resulting in a very low CAH (3.1 $^o$) in the absence of electric voltage. Regardless of the electrode pitch, the apparent CAH under $40$ V is 3 to 6 times larger than the intrinsic CAH. In good agreement with \eq\ref{eq: ewod CAH}, the apparent CAH scales as $\sqrt{p/R}$ for all the droplet volumes and electrode pitches. Although this scaling may seem at odds with well-established theories of CAH based on the Cassie-Baxter framework \cite{choi2009modified}, it is a direct consequence of the assumption of weak contact angle hysteresis yielding a circular droplet. As the CAH becomes larger, the droplet contact line should become trapped over the electrodes \cite{choi2009modified,mchale2011} and follow the better studied regime of contact line motion over strong defects. Using fractal electrodes or electrodes with a broad range of feature sizes would likely yield a more size-independent contact-angle hysteresis. The solid line in Fig. \ref{Hysteresis} shows the model prediction at $40$ V, and confirms the prefactor $\frac{BCU_0^2}{8\gamma_{lg}\sin\theta_E}$. The model begins to underestimates the experimental data at large electrode pitch ($\sqrt{p/R}\simeq 0.25$), after which the droplet starts to lose its circular shape \cite{Vo2020anisotropic, mchale2011} (experimental data for the comparison between contact radii observed in two directions are shown SI).
	

In summary, we report the experimental control of static apparent CAH by an inhomogeneous electric field, in formal analogy with chemical defects. We derive a thermodynamic model to interpret these observations for small CAH. This regime allows considering a droplet circular at all times, and predicts that the CAH depends on the electrode (defect) pitch, which differs from usual CAH studies involving stronger defects. It is also inferred that the CAH grows quadratically with the actuation voltage. These predictions are confirmed against experimental data, even though the model slightly underestimates the experimental hysteresis at large electrode pitch. This study enables the continuous variation of CAH for fundamental studies, and also provides a feasible approach for on-demand and flexible programming of the CAH.

\begin{acknowledgments}
	
	We are indebted to Kaidi Zhang for the fabrication of the first batch of EWOD devices. This work was supported by the National Natural Science Foundation of China with Grant No. 51950410582, 61874033 and 61674043, the Science Foundation of Shanghai Municipal Government with Grant No. 18ZR1402600, and the State Key Lab of ASIC and System, Fudan University with Grant No. 2018MS003.
	
\end{acknowledgments}



\bibliography{reference_2}

\begin{thebibliography}{50}%
\makeatletter
\providecommand \@ifxundefined [1]{%
 \@ifx{#1\undefined}
}%
\providecommand \@ifnum [1]{%
 \ifnum #1\expandafter \@firstoftwo
 \else \expandafter \@secondoftwo
 \fi
}%
\providecommand \@ifx [1]{%
 \ifx #1\expandafter \@firstoftwo
 \else \expandafter \@secondoftwo
 \fi
}%
\providecommand \natexlab [1]{#1}%
\providecommand \enquote  [1]{``#1''}%
\providecommand \bibnamefont  [1]{#1}%
\providecommand \bibfnamefont [1]{#1}%
\providecommand \citenamefont [1]{#1}%
\providecommand \href@noop [0]{\@secondoftwo}%
\providecommand \href [0]{\begingroup \@sanitize@url \@href}%
\providecommand \@href[1]{\@@startlink{#1}\@@href}%
\providecommand \@@href[1]{\endgroup#1\@@endlink}%
\providecommand \@sanitize@url [0]{\catcode `\\12\catcode `\$12\catcode
  `\&12\catcode `\#12\catcode `\^12\catcode `\_12\catcode `\%12\relax}%
\providecommand \@@startlink[1]{}%
\providecommand \@@endlink[0]{}%
\providecommand \url  [0]{\begingroup\@sanitize@url \@url }%
\providecommand \@url [1]{\endgroup\@href {#1}{\urlprefix }}%
\providecommand \urlprefix  [0]{URL }%
\providecommand \Eprint [0]{\href }%
\providecommand \doibase [0]{https://doi.org/}%
\providecommand \selectlanguage [0]{\@gobble}%
\providecommand \bibinfo  [0]{\@secondoftwo}%
\providecommand \bibfield  [0]{\@secondoftwo}%
\providecommand \translation [1]{[#1]}%
\providecommand \BibitemOpen [0]{}%
\providecommand \bibitemStop [0]{}%
\providecommand \bibitemNoStop [0]{.\EOS\space}%
\providecommand \EOS [0]{\spacefactor3000\relax}%
\providecommand \BibitemShut  [1]{\csname bibitem#1\endcsname}%
\let\auto@bib@innerbib\@empty
\bibitem [{\citenamefont {De~Gennes}(1985)}]{Gennes1985}%
  \BibitemOpen
  \bibfield  {author} {\bibinfo {author} {\bibfnamefont {P.~G.}\ \bibnamefont
  {De~Gennes}},\ }\bibfield  {title} {\bibinfo {title} {Wettings - statics and
  dynamics},\ }\href@noop {} {\bibfield  {journal} {\bibinfo  {journal} {Rev.\
  Mod.\ Phys.}\ }\textbf {\bibinfo {volume} {57}},\ \bibinfo {pages} {827}
  (\bibinfo {year} {1985})}\BibitemShut {NoStop}%
\bibitem [{\citenamefont {Eral}\ \emph {et~al.}(2013)\citenamefont {Eral},
  \citenamefont {Mannetje},\ and\ \citenamefont {Oh}}]{Eral2013}%
  \BibitemOpen
  \bibfield  {author} {\bibinfo {author} {\bibfnamefont {H.~B.}\ \bibnamefont
  {Eral}}, \bibinfo {author} {\bibfnamefont {D.~J. C. M.~t.}\ \bibnamefont
  {Mannetje}},\ and\ \bibinfo {author} {\bibfnamefont {J.~M.}\ \bibnamefont
  {Oh}},\ }\bibfield  {title} {\bibinfo {title} {Contact angle hysteresis: a
  review of fundamentals and applications},\ }\href@noop {} {\bibfield
  {journal} {\bibinfo  {journal} {Colloid Polym.\ Sci.}\ }\textbf {\bibinfo
  {volume} {291}},\ \bibinfo {pages} {247} (\bibinfo {year}
  {2013})}\BibitemShut {NoStop}%
\bibitem [{\citenamefont {Erbil}(2014)}]{Erbil2014}%
  \BibitemOpen
  \bibfield  {author} {\bibinfo {author} {\bibfnamefont {H.~Y.}\ \bibnamefont
  {Erbil}},\ }\bibfield  {title} {\bibinfo {title} {The debate on the
  dependence of apparent contact angles on drop contact area or three-phase
  contact line: A review},\ }\href@noop {} {\bibfield  {journal} {\bibinfo
  {journal} {Surf.\ Sci.\ Rep.}\ }\textbf {\bibinfo {volume} {69}},\ \bibinfo
  {pages} {325} (\bibinfo {year} {2014})}\BibitemShut {NoStop}%
\bibitem [{\citenamefont {Johnson}\ and\ \citenamefont
  {Dettre}(1964{\natexlab{a}})}]{Johnson1964}%
  \BibitemOpen
  \bibfield  {author} {\bibinfo {author} {\bibfnamefont {R.~E.~J.}\
  \bibnamefont {Johnson}}\ and\ \bibinfo {author} {\bibfnamefont {R.~H.}\
  \bibnamefont {Dettre}},\ }\bibfield  {title} {\bibinfo {title} {Contact angle
  hysteresis i. study of an idealized rough surface},\ }in\ \href@noop {}
  {\emph {\bibinfo {booktitle} {Contact Angle, Wettability, and Adhesion}}},\
  \bibinfo {series and number} {Advances in Chemistry}\ (\bibinfo  {publisher}
  {American Chemical Society},\ \bibinfo {year} {1964})\ Chap.~\bibinfo
  {chapter} {7}, pp.\ \bibinfo {pages} {112--135}\BibitemShut {NoStop}%
\bibitem [{\citenamefont {Cox}(1983)}]{cox1983spreading}%
  \BibitemOpen
  \bibfield  {author} {\bibinfo {author} {\bibfnamefont {R.~G.}\ \bibnamefont
  {Cox}},\ }\bibfield  {title} {\bibinfo {title} {The spreading of a liquid on
  a rough solid surface},\ }\href@noop {} {\bibfield  {journal} {\bibinfo
  {journal} {J.\ Fluid Mech.}\ }\textbf {\bibinfo {volume} {131}},\ \bibinfo
  {pages} {1} (\bibinfo {year} {1983})}\BibitemShut {NoStop}%
\bibitem [{\citenamefont {Johnson}\ and\ \citenamefont
  {Dettre}(1964{\natexlab{b}})}]{Johnson1964b}%
  \BibitemOpen
  \bibfield  {author} {\bibinfo {author} {\bibfnamefont {R.~E.~J.}\
  \bibnamefont {Johnson}}\ and\ \bibinfo {author} {\bibfnamefont {R.~H.}\
  \bibnamefont {Dettre}},\ }\bibfield  {title} {\bibinfo {title} {Contact angle
  hysteresis. iii. study of an idealized heterogeneous surface},\ }\href@noop
  {} {\bibfield  {journal} {\bibinfo  {journal} {\ J.\ Phys.\ Chem.}\ }\textbf
  {\bibinfo {volume} {68}},\ \bibinfo {pages} {1744} (\bibinfo {year}
  {1964}{\natexlab{b}})}\BibitemShut {NoStop}%
\bibitem [{\citenamefont {Priest}\ \emph {et~al.}(2007)\citenamefont {Priest},
  \citenamefont {Sedev},\ and\ \citenamefont {Ralston}}]{Priest2007}%
  \BibitemOpen
  \bibfield  {author} {\bibinfo {author} {\bibfnamefont {C.}~\bibnamefont
  {Priest}}, \bibinfo {author} {\bibfnamefont {R.}~\bibnamefont {Sedev}},\ and\
  \bibinfo {author} {\bibfnamefont {J.}~\bibnamefont {Ralston}},\ }\bibfield
  {title} {\bibinfo {title} {Asymmetric wetting hysteresis on chemical
  defects},\ }\href@noop {} {\bibfield  {journal} {\bibinfo  {journal} {Phys.\
  Rev.\ Lett.}\ }\textbf {\bibinfo {volume} {99}},\ \bibinfo {pages} {026103}
  (\bibinfo {year} {2007})}\BibitemShut {NoStop}%
\bibitem [{\citenamefont {De~Johghe}\ and\ \citenamefont
  {Chatain}(1995)}]{chatain1995experimental}%
  \BibitemOpen
  \bibfield  {author} {\bibinfo {author} {\bibfnamefont {V.}~\bibnamefont
  {De~Johghe}}\ and\ \bibinfo {author} {\bibfnamefont {D.}~\bibnamefont
  {Chatain}},\ }\bibfield  {title} {\bibinfo {title} {Experimental study of
  wetting hysteresis on surfaces with controlled geometrical and/or chemical
  defects},\ }\href@noop {} {\bibfield  {journal} {\bibinfo  {journal} {Acta
  metall. mater.}\ }\textbf {\bibinfo {volume} {43}},\ \bibinfo {pages} {1505}
  (\bibinfo {year} {1995})}\BibitemShut {NoStop}%
\bibitem [{\citenamefont {Good}(1952)}]{Good1952}%
  \BibitemOpen
  \bibfield  {author} {\bibinfo {author} {\bibfnamefont {R.~J.}\ \bibnamefont
  {Good}},\ }\bibfield  {title} {\bibinfo {title} {A thermodynamic derviation
  of \ wenzel's modification of \ young's equation for contact angles -
  together with a theory of hysteresis},\ }\href
  {https://doi.org/10.1021/ja01140a014} {\bibfield  {journal} {\bibinfo
  {journal} {J.\ Am \ Chem.\ Soc.}\ }\textbf {\bibinfo {volume} {74}},\
  \bibinfo {pages} {5041} (\bibinfo {year} {1952})}\BibitemShut {NoStop}%
\bibitem [{\citenamefont {Sun}\ \emph {et~al.}(2019)\citenamefont {Sun},
  \citenamefont {Wang}, \citenamefont {Li}, \citenamefont {Zhang},
  \citenamefont {Ye}, \citenamefont {Cui}, \citenamefont {Chen}, \citenamefont
  {Wang}, \citenamefont {Butt}, \citenamefont {Vollmer},\ and\ \citenamefont
  {Deng}}]{sun2019surface}%
  \BibitemOpen
  \bibfield  {author} {\bibinfo {author} {\bibfnamefont {Q.}~\bibnamefont
  {Sun}}, \bibinfo {author} {\bibfnamefont {D.}~\bibnamefont {Wang}}, \bibinfo
  {author} {\bibfnamefont {Y.}~\bibnamefont {Li}}, \bibinfo {author}
  {\bibfnamefont {J.}~\bibnamefont {Zhang}}, \bibinfo {author} {\bibfnamefont
  {S.}~\bibnamefont {Ye}}, \bibinfo {author} {\bibfnamefont {J.}~\bibnamefont
  {Cui}}, \bibinfo {author} {\bibfnamefont {L.}~\bibnamefont {Chen}}, \bibinfo
  {author} {\bibfnamefont {Z.}~\bibnamefont {Wang}}, \bibinfo {author}
  {\bibfnamefont {H.-J.}\ \bibnamefont {Butt}}, \bibinfo {author}
  {\bibfnamefont {D.}~\bibnamefont {Vollmer}},\ and\ \bibinfo {author}
  {\bibfnamefont {X.}~\bibnamefont {Deng}},\ }\bibfield  {title} {\bibinfo
  {title} {Surface charge printing for programmed droplet transport},\
  }\href@noop {} {\bibfield  {journal} {\bibinfo  {journal} {Nat. Mater.}\
  }\textbf {\bibinfo {volume} {18}},\ \bibinfo {pages} {936} (\bibinfo {year}
  {2019})}\BibitemShut {NoStop}%
\bibitem [{\citenamefont {Mugele}\ and\ \citenamefont
  {Baret}(2005)}]{mugele2005electrowetting}%
  \BibitemOpen
  \bibfield  {author} {\bibinfo {author} {\bibfnamefont {F.}~\bibnamefont
  {Mugele}}\ and\ \bibinfo {author} {\bibfnamefont {J.~C.}\ \bibnamefont
  {Baret}},\ }\bibfield  {title} {\bibinfo {title} {Electrowetting: from basics
  to applications},\ }\href@noop {} {\bibfield  {journal} {\bibinfo  {journal}
  {J.\ Phys.\: Condens.\ Matter}\ }\textbf {\bibinfo {volume} {17}},\ \bibinfo
  {pages} {R705} (\bibinfo {year} {2005})}\BibitemShut {NoStop}%
\bibitem [{\citenamefont {Fair}(2007)}]{Fair2007}%
  \BibitemOpen
  \bibfield  {author} {\bibinfo {author} {\bibfnamefont {R.~B.}\ \bibnamefont
  {Fair}},\ }\bibfield  {title} {\bibinfo {title} {Digital microfluidics: is a
  true lab-on-a-chip possible?},\ }\href@noop {} {\bibfield  {journal}
  {\bibinfo  {journal} {Microfluid.\ Nanofluid.}\ }\textbf {\bibinfo {volume}
  {3}},\ \bibinfo {pages} {245} (\bibinfo {year} {2007})}\BibitemShut {NoStop}%
\bibitem [{\citenamefont {Nelson}\ and\ \citenamefont
  {Kim}(2012)}]{Nelson2012}%
  \BibitemOpen
  \bibfield  {author} {\bibinfo {author} {\bibfnamefont {W.}~\bibnamefont
  {Nelson}}\ and\ \bibinfo {author} {\bibfnamefont {C.~J.}\ \bibnamefont
  {Kim}},\ }\bibfield  {title} {\bibinfo {title} {Droplet actuation by
  electrowetting-on-dielectric (ewod): A review},\ }\href@noop {} {\bibfield
  {journal} {\bibinfo  {journal} {J.\ Adhes.\ Sci.\ Technol.}\ }\textbf
  {\bibinfo {volume} {26}},\ \bibinfo {pages} {1747} (\bibinfo {year}
  {2012})}\BibitemShut {NoStop}%
\bibitem [{\citenamefont {Heikenfeld}\ \emph {et~al.}(2009)\citenamefont
  {Heikenfeld}, \citenamefont {Zhou}, \citenamefont {Kreit}, \citenamefont
  {Raj}, \citenamefont {Yang}, \citenamefont {Sun}, \citenamefont {Milarcik},
  \citenamefont {Clapp},\ and\ \citenamefont {Schwartz}}]{Heikenfeld2009}%
  \BibitemOpen
  \bibfield  {author} {\bibinfo {author} {\bibfnamefont {J.}~\bibnamefont
  {Heikenfeld}}, \bibinfo {author} {\bibfnamefont {K.}~\bibnamefont {Zhou}},
  \bibinfo {author} {\bibfnamefont {E.}~\bibnamefont {Kreit}}, \bibinfo
  {author} {\bibfnamefont {B.}~\bibnamefont {Raj}}, \bibinfo {author}
  {\bibfnamefont {S.}~\bibnamefont {Yang}}, \bibinfo {author} {\bibfnamefont
  {B.}~\bibnamefont {Sun}}, \bibinfo {author} {\bibfnamefont {A.}~\bibnamefont
  {Milarcik}}, \bibinfo {author} {\bibfnamefont {L.}~\bibnamefont {Clapp}},\
  and\ \bibinfo {author} {\bibfnamefont {R.}~\bibnamefont {Schwartz}},\
  }\bibfield  {title} {\bibinfo {title} {Electrofluidic displays using
  young-laplace transposition of brilliant pigment dispersions},\ }\href@noop
  {} {\bibfield  {journal} {\bibinfo  {journal} {Nat.\ Photonics}\ }\textbf
  {\bibinfo {volume} {3}},\ \bibinfo {pages} {292} (\bibinfo {year}
  {2009})}\BibitemShut {NoStop}%
\bibitem [{\citenamefont {Hayes}\ and\ \citenamefont
  {Feenstra}(2003)}]{Hayes2003}%
  \BibitemOpen
  \bibfield  {author} {\bibinfo {author} {\bibfnamefont {R.~A.}\ \bibnamefont
  {Hayes}}\ and\ \bibinfo {author} {\bibfnamefont {B.~J.}\ \bibnamefont
  {Feenstra}},\ }\bibfield  {title} {\bibinfo {title} {Video-speed electronic
  paper based on electrowetting},\ }\href@noop {} {\bibfield  {journal}
  {\bibinfo  {journal} {Nature}\ }\textbf {\bibinfo {volume} {425}},\ \bibinfo
  {pages} {383} (\bibinfo {year} {2003})}\BibitemShut {NoStop}%
\bibitem [{\citenamefont {Li}\ and\ \citenamefont {Jiang}(2012)}]{Li2012}%
  \BibitemOpen
  \bibfield  {author} {\bibinfo {author} {\bibfnamefont {C.}~\bibnamefont
  {Li}}\ and\ \bibinfo {author} {\bibfnamefont {H.}~\bibnamefont {Jiang}},\
  }\bibfield  {title} {\bibinfo {title} {Electrowetting-driven variable-focus
  microlens on flexible surfaces},\ }\href@noop {} {\bibfield  {journal}
  {\bibinfo  {journal} {Appl.\ Phys.\ Lett.}\ }\textbf {\bibinfo {volume}
  {100}},\ \bibinfo {pages} {231105} (\bibinfo {year} {2012})}\BibitemShut
  {NoStop}%
\bibitem [{\citenamefont {Hao}\ \emph {et~al.}(2014)\citenamefont {Hao},
  \citenamefont {Liu}, \citenamefont {Chen}, \citenamefont {He}, \citenamefont
  {Li}, \citenamefont {Li},\ and\ \citenamefont {Wang}}]{Hao2014}%
  \BibitemOpen
  \bibfield  {author} {\bibinfo {author} {\bibfnamefont {C.}~\bibnamefont
  {Hao}}, \bibinfo {author} {\bibfnamefont {Y.}~\bibnamefont {Liu}}, \bibinfo
  {author} {\bibfnamefont {X.}~\bibnamefont {Chen}}, \bibinfo {author}
  {\bibfnamefont {Y.}~\bibnamefont {He}}, \bibinfo {author} {\bibfnamefont
  {Q.}~\bibnamefont {Li}}, \bibinfo {author} {\bibfnamefont {K.~Y.}\
  \bibnamefont {Li}},\ and\ \bibinfo {author} {\bibfnamefont {Z.}~\bibnamefont
  {Wang}},\ }\bibfield  {title} {\bibinfo {title} {Electrowetting on
  liquid-infused film (ewolf): complete reversibility and controlled droplet
  oscillation suppression for fast optical imaging},\ }\href@noop {} {\bibfield
   {journal} {\bibinfo  {journal} {Sci.\ Rep.}\ }\textbf {\bibinfo {volume}
  {4}},\ \bibinfo {pages} {6846} (\bibinfo {year} {2014})}\BibitemShut
  {NoStop}%
\bibitem [{\citenamefont {Heikenfeld}(2005)}]{Heikenfeld2005}%
  \BibitemOpen
  \bibfield  {author} {\bibinfo {author} {\bibfnamefont {J.}~\bibnamefont
  {Heikenfeld}},\ }\bibfield  {title} {\bibinfo {title} {High-transmission
  electrowetting light valves},\ }\href@noop {} {\bibfield  {journal} {\bibinfo
   {journal} {Appl.\ Phys.\ Lett.}\ }\textbf {\bibinfo {volume} {86}},\
  \bibinfo {pages} {151121} (\bibinfo {year} {2005})}\BibitemShut {NoStop}%
\bibitem [{\citenamefont {Murade}\ \emph {et~al.}(2011)\citenamefont {Murade},
  \citenamefont {Oh}, \citenamefont {van~den Ende},\ and\ \citenamefont
  {Muggle}}]{Murade2011}%
  \BibitemOpen
  \bibfield  {author} {\bibinfo {author} {\bibfnamefont {C.~U.}\ \bibnamefont
  {Murade}}, \bibinfo {author} {\bibfnamefont {J.~M.}\ \bibnamefont {Oh}},
  \bibinfo {author} {\bibfnamefont {D.}~\bibnamefont {van~den Ende}},\ and\
  \bibinfo {author} {\bibfnamefont {F.}~\bibnamefont {Muggle}},\ }\bibfield
  {title} {\bibinfo {title} {Electrowetting driven optical switch and tunable
  aperture},\ }\href@noop {} {\bibfield  {journal} {\bibinfo  {journal} {Opt.\
  Express}\ }\textbf {\bibinfo {volume} {19}},\ \bibinfo {pages} {15525}
  (\bibinfo {year} {2011})}\BibitemShut {NoStop}%
\bibitem [{\citenamefont {Lippmann}(1875)}]{Lippmann1875}%
  \BibitemOpen
  \bibfield  {author} {\bibinfo {author} {\bibfnamefont {M.~G.}\ \bibnamefont
  {Lippmann}},\ }\bibfield  {title} {\bibinfo {title} {Relation entre les
  phenomenes electriques et capillaires},\ }\href@noop {} {\bibfield  {journal}
  {\bibinfo  {journal} {Ann.\ Chim.\ Phys.}\ }\textbf {\bibinfo {volume} {5}},\
  \bibinfo {pages} {494} (\bibinfo {year} {1875})}\BibitemShut {NoStop}%
\bibitem [{\citenamefont {Jones}(2002)}]{jones2002relationship}%
  \BibitemOpen
  \bibfield  {author} {\bibinfo {author} {\bibfnamefont {T.~B.}\ \bibnamefont
  {Jones}},\ }\bibfield  {title} {\bibinfo {title} {On the relationship of
  dielectrophoresis and electrowetting},\ }\href@noop {} {\bibfield  {journal}
  {\bibinfo  {journal} {Langmuir}\ }\textbf {\bibinfo {volume} {18}},\ \bibinfo
  {pages} {4437} (\bibinfo {year} {2002})}\BibitemShut {NoStop}%
\bibitem [{\citenamefont {Buehrle}\ \emph {et~al.}(2003)\citenamefont
  {Buehrle}, \citenamefont {Herminghaus},\ and\ \citenamefont
  {Mugele}}]{Buehrle2003}%
  \BibitemOpen
  \bibfield  {author} {\bibinfo {author} {\bibfnamefont {J.}~\bibnamefont
  {Buehrle}}, \bibinfo {author} {\bibfnamefont {S.}~\bibnamefont
  {Herminghaus}},\ and\ \bibinfo {author} {\bibfnamefont {F.}~\bibnamefont
  {Mugele}},\ }\bibfield  {title} {\bibinfo {title} {Interface profiles near
  three-phase contact lines in electric fields},\ }\href@noop {} {\bibfield
  {journal} {\bibinfo  {journal} {Phys.\ Rev.\ Lett.}\ }\textbf {\bibinfo
  {volume} {91}},\ \bibinfo {pages} {086101} (\bibinfo {year}
  {2003})}\BibitemShut {NoStop}%
\bibitem [{\citenamefont {Bonn}\ \emph {et~al.}(2009)\citenamefont {Bonn},
  \citenamefont {Eggers}, \citenamefont {Indekeu}, \citenamefont {Meunier},\
  and\ \citenamefont {Rolley}}]{Bonn2009}%
  \BibitemOpen
  \bibfield  {author} {\bibinfo {author} {\bibfnamefont {D.}~\bibnamefont
  {Bonn}}, \bibinfo {author} {\bibfnamefont {J.}~\bibnamefont {Eggers}},
  \bibinfo {author} {\bibfnamefont {J.}~\bibnamefont {Indekeu}}, \bibinfo
  {author} {\bibfnamefont {J.}~\bibnamefont {Meunier}},\ and\ \bibinfo {author}
  {\bibfnamefont {E.}~\bibnamefont {Rolley}},\ }\bibfield  {title} {\bibinfo
  {title} {Wetting and spreading},\ }\href@noop {} {\bibfield  {journal}
  {\bibinfo  {journal} {Rev.\ Mod.\ Phys.}\ }\textbf {\bibinfo {volume} {81}},\
  \bibinfo {pages} {739} (\bibinfo {year} {2009})}\BibitemShut {NoStop}%
\bibitem [{\citenamefont {Mugele}\ and\ \citenamefont
  {Buehrle}(2007)}]{mugele2007equilibrium}%
  \BibitemOpen
  \bibfield  {author} {\bibinfo {author} {\bibfnamefont {F.}~\bibnamefont
  {Mugele}}\ and\ \bibinfo {author} {\bibfnamefont {J.}~\bibnamefont
  {Buehrle}},\ }\bibfield  {title} {\bibinfo {title} {Equilibrium drop surface
  profiles in electric fields},\ }\href@noop {} {\bibfield  {journal} {\bibinfo
   {journal} {J. Phys.: Condens. Matter}\ }\textbf {\bibinfo {volume} {19}},\
  \bibinfo {pages} {375112} (\bibinfo {year} {2007})}\BibitemShut {NoStop}%
\bibitem [{\citenamefont
  {Israelachvili}(2011)}]{israelachvili2011intermolecular}%
  \BibitemOpen
  \bibinfo {editor} {\bibfnamefont {J.~N.}\ \bibnamefont {Israelachvili}},\
  ed.,\ in\ \href
  {https://doi.org/https://doi.org/10.1016/B978-0-12-375182-9.10024-7} {\emph
  {\bibinfo {booktitle} {Intermolecular and Surface Forces (Third Edition)}}}\
  (\bibinfo  {publisher} {Academic Press},\ \bibinfo {address} {San Diego},\
  \bibinfo {year} {2011})\ Chap.\ \bibinfo {chapter} {17.1}, p.\ \bibinfo
  {pages} {415},\ \bibinfo {edition} {third edition}\ ed.\BibitemShut {Stop}%
\bibitem [{\citenamefont {Nelson}\ \emph {et~al.}(2011)\citenamefont {Nelson},
  \citenamefont {Sen},\ and\ \citenamefont {Kim}}]{nelson2011dynamic}%
  \BibitemOpen
  \bibfield  {author} {\bibinfo {author} {\bibfnamefont {W.~C.}\ \bibnamefont
  {Nelson}}, \bibinfo {author} {\bibfnamefont {P.}~\bibnamefont {Sen}},\ and\
  \bibinfo {author} {\bibfnamefont {C.~J.}\ \bibnamefont {Kim}},\ }\bibfield
  {title} {\bibinfo {title} {Dynamic contact angles and hysteresis under
  electrowetting-on-dielectric},\ }\href@noop {} {\bibfield  {journal}
  {\bibinfo  {journal} {Langmuir}\ }\textbf {\bibinfo {volume} {27}},\ \bibinfo
  {pages} {10319} (\bibinfo {year} {2011})}\BibitemShut {NoStop}%
\bibitem [{\citenamefont {McHale}\ \emph {et~al.}(2011)\citenamefont {McHale},
  \citenamefont {Brown}, \citenamefont {Newton}, \citenamefont {Wells},\ and\
  \citenamefont {Sampara}}]{mchale2011}%
  \BibitemOpen
  \bibfield  {author} {\bibinfo {author} {\bibfnamefont {G.}~\bibnamefont
  {McHale}}, \bibinfo {author} {\bibfnamefont {C.~V.}\ \bibnamefont {Brown}},
  \bibinfo {author} {\bibfnamefont {M.~I.}\ \bibnamefont {Newton}}, \bibinfo
  {author} {\bibfnamefont {G.~G.}\ \bibnamefont {Wells}},\ and\ \bibinfo
  {author} {\bibfnamefont {N.}~\bibnamefont {Sampara}},\ }\bibfield  {title}
  {\bibinfo {title} {Dielectrowetting driven spreading of droplets},\
  }\href@noop {} {\bibfield  {journal} {\bibinfo  {journal} {Phys.\ Rev.\
  Lett.}\ }\textbf {\bibinfo {volume} {107}},\ \bibinfo {pages} {186101}
  (\bibinfo {year} {2011})}\BibitemShut {NoStop}%
\bibitem [{\citenamefont {Bhushan}\ and\ \citenamefont
  {Pan}(2011)}]{bhushan2011role}%
  \BibitemOpen
  \bibfield  {author} {\bibinfo {author} {\bibfnamefont {B.}~\bibnamefont
  {Bhushan}}\ and\ \bibinfo {author} {\bibfnamefont {Y.}~\bibnamefont {Pan}},\
  }\bibfield  {title} {\bibinfo {title} {Role of electric field on surface
  wetting of polystyrene surface},\ }\href@noop {} {\bibfield  {journal}
  {\bibinfo  {journal} {Langmuir}\ }\textbf {\bibinfo {volume} {27}},\ \bibinfo
  {pages} {9425} (\bibinfo {year} {2011})}\BibitemShut {NoStop}%
\bibitem [{\citenamefont {Li}\ and\ \citenamefont {Mugele}(2008)}]{li2008make}%
  \BibitemOpen
  \bibfield  {author} {\bibinfo {author} {\bibfnamefont {F.}~\bibnamefont
  {Li}}\ and\ \bibinfo {author} {\bibfnamefont {F.}~\bibnamefont {Mugele}},\
  }\bibfield  {title} {\bibinfo {title} {How to make sticky surfaces slippery:
  Contact angle hysteresis in electrowetting with alternating voltage},\
  }\href@noop {} {\bibfield  {journal} {\bibinfo  {journal} {Appl.\ Phys.\
  Lett.}\ }\textbf {\bibinfo {volume} {92}},\ \bibinfo {pages} {244108}
  (\bibinfo {year} {2008})}\BibitemShut {NoStop}%
\bibitem [{\citenamefont {Nita}\ \emph {et~al.}(2017)\citenamefont {Nita},
  \citenamefont {Do-Quang}, \citenamefont {Wang}, \citenamefont {Chen},
  \citenamefont {Suzuki}, \citenamefont {Amberg},\ and\ \citenamefont
  {Shiomi}}]{nita2017electrostatic}%
  \BibitemOpen
  \bibfield  {author} {\bibinfo {author} {\bibfnamefont {S.}~\bibnamefont
  {Nita}}, \bibinfo {author} {\bibfnamefont {M.}~\bibnamefont {Do-Quang}},
  \bibinfo {author} {\bibfnamefont {J.}~\bibnamefont {Wang}}, \bibinfo {author}
  {\bibfnamefont {Y.}~\bibnamefont {Chen}}, \bibinfo {author} {\bibfnamefont
  {Y.}~\bibnamefont {Suzuki}}, \bibinfo {author} {\bibfnamefont
  {G.}~\bibnamefont {Amberg}},\ and\ \bibinfo {author} {\bibfnamefont
  {J.}~\bibnamefont {Shiomi}},\ }\bibfield  {title} {\bibinfo {title}
  {Electrostatic cloaking of surface structure for dynamic wetting},\
  }\href@noop {} {\bibfield  {journal} {\bibinfo  {journal} {Sci. Adv.}\
  }\textbf {\bibinfo {volume} {3}},\ \bibinfo {pages} {e1602202} (\bibinfo
  {year} {2017})}\BibitemShut {NoStop}%
\bibitem [{\citenamefont {Yi}\ and\ \citenamefont {Kim}(2006)}]{Yi2006}%
  \BibitemOpen
  \bibfield  {author} {\bibinfo {author} {\bibfnamefont {U.~C.}\ \bibnamefont
  {Yi}}\ and\ \bibinfo {author} {\bibfnamefont {C.~J.}\ \bibnamefont {Kim}},\
  }\bibfield  {title} {\bibinfo {title} {Characterization of electrowetting
  actuation on addressable single-side coplanar electrodes},\ }\href@noop {}
  {\bibfield  {journal} {\bibinfo  {journal} {J.\ Micromech.\ Microeng.}\
  }\textbf {\bibinfo {volume} {16}},\ \bibinfo {pages} {2053} (\bibinfo {year}
  {2006})}\BibitemShut {NoStop}%
\bibitem [{\citenamefont {Wang}\ and\ \citenamefont
  {Bhushan}(2009)}]{wang2009liquid}%
  \BibitemOpen
  \bibfield  {author} {\bibinfo {author} {\bibfnamefont {Y.}~\bibnamefont
  {Wang}}\ and\ \bibinfo {author} {\bibfnamefont {B.}~\bibnamefont {Bhushan}},\
  }\bibfield  {title} {\bibinfo {title} {Liquid microdroplet sliding on
  hydrophobic surfaces in the presence of an electric field},\ }\href@noop {}
  {\bibfield  {journal} {\bibinfo  {journal} {Langmuir}\ }\textbf {\bibinfo
  {volume} {26}},\ \bibinfo {pages} {4013} (\bibinfo {year}
  {2009})}\BibitemShut {NoStop}%
\bibitem [{\citenamefont {Manukyan}\ \emph {et~al.}(2011)\citenamefont
  {Manukyan}, \citenamefont {Oh}, \citenamefont {van~den Ende}, \citenamefont
  {Lammertink},\ and\ \citenamefont {Mugele}}]{manukyan2011electrical}%
  \BibitemOpen
  \bibfield  {author} {\bibinfo {author} {\bibfnamefont {G.}~\bibnamefont
  {Manukyan}}, \bibinfo {author} {\bibfnamefont {J.~M.}\ \bibnamefont {Oh}},
  \bibinfo {author} {\bibfnamefont {D.}~\bibnamefont {van~den Ende}}, \bibinfo
  {author} {\bibfnamefont {R.~G.~H.}\ \bibnamefont {Lammertink}},\ and\
  \bibinfo {author} {\bibfnamefont {F.}~\bibnamefont {Mugele}},\ }\bibfield
  {title} {\bibinfo {title} {Electrical switching of wetting states on
  superhydrophobic surfaces: A route towards reversible cassie-to-wenzel
  transitions},\ }\href@noop {} {\bibfield  {journal} {\bibinfo  {journal}
  {Phys.\ Rev.\ Lett.}\ }\textbf {\bibinfo {volume} {106}},\ \bibinfo {pages}
  {014501} (\bibinfo {year} {2011})}\BibitemShut {NoStop}%
\bibitem [{\citenamefont {Morgan}\ \emph {et~al.}(2001)\citenamefont {Morgan},
  \citenamefont {Izquierdo}, \citenamefont {Bakewell}, \citenamefont {Green},\
  and\ \citenamefont {Ramos}}]{Morgen2001}%
  \BibitemOpen
  \bibfield  {author} {\bibinfo {author} {\bibfnamefont {H.}~\bibnamefont
  {Morgan}}, \bibinfo {author} {\bibfnamefont {A.~G.}\ \bibnamefont
  {Izquierdo}}, \bibinfo {author} {\bibfnamefont {D.}~\bibnamefont {Bakewell}},
  \bibinfo {author} {\bibfnamefont {N.~G.}\ \bibnamefont {Green}},\ and\
  \bibinfo {author} {\bibfnamefont {A.}~\bibnamefont {Ramos}},\ }\bibfield
  {title} {\bibinfo {title} {The dielectrophoretic and travelling wave forces
  generated by interdigitated electrode arrays: analytical solution using
  fourier series},\ }\href@noop {} {\bibfield  {journal} {\bibinfo  {journal}
  {J.\ Phys.\ D: Appl.\ Phys.}\ }\textbf {\bibinfo {volume} {34}},\ \bibinfo
  {pages} {1553} (\bibinfo {year} {2001})}\BibitemShut {NoStop}%
\bibitem [{\citenamefont {Gunda}\ and\ \citenamefont
  {Mitra}(2010)}]{Gunda2010}%
  \BibitemOpen
  \bibfield  {author} {\bibinfo {author} {\bibfnamefont {N.~S.~K.}\
  \bibnamefont {Gunda}}\ and\ \bibinfo {author} {\bibfnamefont {S.~K.}\
  \bibnamefont {Mitra}},\ }\bibfield  {title} {\bibinfo {title} {Modeling of
  dielectrophoretic transport of myoglobin molecules in microchannels},\
  }\href@noop {} {\bibfield  {journal} {\bibinfo  {journal} {Biomicrofluidics}\
  }\textbf {\bibinfo {volume} {4}},\ \bibinfo {pages} {014105} (\bibinfo {year}
  {2010})}\BibitemShut {NoStop}%
\bibitem [{\citenamefont {Joanny}\ and\ \citenamefont
  {De~Gennes}(1984)}]{Joanny1984}%
  \BibitemOpen
  \bibfield  {author} {\bibinfo {author} {\bibfnamefont {J.~F.}\ \bibnamefont
  {Joanny}}\ and\ \bibinfo {author} {\bibfnamefont {P.~G.}\ \bibnamefont
  {De~Gennes}},\ }\bibfield  {title} {\bibinfo {title} {A model for contact
  angle hysteresis},\ }\href@noop {} {\bibfield  {journal} {\bibinfo  {journal}
  {J.\ Chem.\ Phys.}\ }\textbf {\bibinfo {volume} {81}},\ \bibinfo {pages}
  {552} (\bibinfo {year} {1984})}\BibitemShut {NoStop}%
\bibitem [{\citenamefont {Yuan}\ and\ \citenamefont
  {Zhao}(2010)}]{yuan2010precursor}%
  \BibitemOpen
  \bibfield  {author} {\bibinfo {author} {\bibfnamefont {Q.}~\bibnamefont
  {Yuan}}\ and\ \bibinfo {author} {\bibfnamefont {Y.~P.}\ \bibnamefont
  {Zhao}},\ }\bibfield  {title} {\bibinfo {title} {Precursor film in dynamic
  wetting, electrowetting, and electro-elasto-capillarity},\ }\href@noop {}
  {\bibfield  {journal} {\bibinfo  {journal} {Phys.\ Rev.\ Lett.}\ }\textbf
  {\bibinfo {volume} {104}},\ \bibinfo {pages} {246101} (\bibinfo {year}
  {2010})}\BibitemShut {NoStop}%
\bibitem [{\citenamefont {Huh}\ and\ \citenamefont
  {Mason}(1977)}]{huh1977effects}%
  \BibitemOpen
  \bibfield  {author} {\bibinfo {author} {\bibfnamefont {C.}~\bibnamefont
  {Huh}}\ and\ \bibinfo {author} {\bibfnamefont {S.~G.}\ \bibnamefont
  {Mason}},\ }\bibfield  {title} {\bibinfo {title} {Effects of surface
  roughness on wetting (theoretical)},\ }\href@noop {} {\bibfield  {journal}
  {\bibinfo  {journal} {J.\ colloid Interf.\ Sci.}\ }\textbf {\bibinfo {volume}
  {60}},\ \bibinfo {pages} {11} (\bibinfo {year} {1977})}\BibitemShut {NoStop}%
\bibitem [{\citenamefont {Wang}\ \emph {et~al.}(2020)\citenamefont {Wang},
  \citenamefont {Wang}, \citenamefont {Zhang}, \citenamefont {Wang},
  \citenamefont {Riaud},\ and\ \citenamefont {Zhou}}]{wang2020demand}%
  \BibitemOpen
  \bibfield  {author} {\bibinfo {author} {\bibfnamefont {W.}~\bibnamefont
  {Wang}}, \bibinfo {author} {\bibfnamefont {Q.}~\bibnamefont {Wang}}, \bibinfo
  {author} {\bibfnamefont {K.}~\bibnamefont {Zhang}}, \bibinfo {author}
  {\bibfnamefont {X.}~\bibnamefont {Wang}}, \bibinfo {author} {\bibfnamefont
  {A.}~\bibnamefont {Riaud}},\ and\ \bibinfo {author} {\bibfnamefont
  {J.}~\bibnamefont {Zhou}},\ }\bibfield  {title} {\bibinfo {title} {On-demand
  contact line pinning during droplet evaporation},\ }\href@noop {} {\bibfield
  {journal} {\bibinfo  {journal} {Sens. Actuators, B}\ ,\ \bibinfo {pages}
  {127983}} (\bibinfo {year} {2020})}\BibitemShut {NoStop}%
\bibitem [{\citenamefont {Choi}\ \emph {et~al.}(2009)\citenamefont {Choi},
  \citenamefont {Tuteja}, \citenamefont {Mabry}, \citenamefont {Cohen},\ and\
  \citenamefont {McKinley}}]{choi2009modified}%
  \BibitemOpen
  \bibfield  {author} {\bibinfo {author} {\bibfnamefont {W.}~\bibnamefont
  {Choi}}, \bibinfo {author} {\bibfnamefont {A.}~\bibnamefont {Tuteja}},
  \bibinfo {author} {\bibfnamefont {J.~M.}\ \bibnamefont {Mabry}}, \bibinfo
  {author} {\bibfnamefont {R.~E.}\ \bibnamefont {Cohen}},\ and\ \bibinfo
  {author} {\bibfnamefont {G.~H.}\ \bibnamefont {McKinley}},\ }\bibfield
  {title} {\bibinfo {title} {A modified cassie--baxter relationship to explain
  contact angle hysteresis and anisotropy on non-wetting textured surfaces},\
  }\href@noop {} {\bibfield  {journal} {\bibinfo  {journal} {J. Colloid
  Interface Sci.}\ }\textbf {\bibinfo {volume} {339}},\ \bibinfo {pages} {208}
  (\bibinfo {year} {2009})}\BibitemShut {NoStop}%
\bibitem [{\citenamefont {Chen}\ \emph {et~al.}(2014)\citenamefont {Chen},
  \citenamefont {Yu}, \citenamefont {Zhang}, \citenamefont {Wu}, \citenamefont
  {Liu},\ and\ \citenamefont {Zhou}}]{chen2014study}%
  \BibitemOpen
  \bibfield  {author} {\bibinfo {author} {\bibfnamefont {J.}~\bibnamefont
  {Chen}}, \bibinfo {author} {\bibfnamefont {Y.}~\bibnamefont {Yu}}, \bibinfo
  {author} {\bibfnamefont {K.}~\bibnamefont {Zhang}}, \bibinfo {author}
  {\bibfnamefont {C.}~\bibnamefont {Wu}}, \bibinfo {author} {\bibfnamefont
  {A.~Q.}\ \bibnamefont {Liu}},\ and\ \bibinfo {author} {\bibfnamefont
  {J.}~\bibnamefont {Zhou}},\ }\bibfield  {title} {\bibinfo {title} {Study of
  cyanoethyl pullulan as insulator for electrowetting},\ }\href@noop {}
  {\bibfield  {journal} {\bibinfo  {journal} {Sens. Actuators, B}\ }\textbf
  {\bibinfo {volume} {199}},\ \bibinfo {pages} {183} (\bibinfo {year}
  {2014})}\BibitemShut {NoStop}%
\bibitem [{\citenamefont {Mugele}\ and\ \citenamefont
  {Herminghaus}(2002)}]{Mugele2002}%
  \BibitemOpen
  \bibfield  {author} {\bibinfo {author} {\bibfnamefont {F.}~\bibnamefont
  {Mugele}}\ and\ \bibinfo {author} {\bibfnamefont {S.}~\bibnamefont
  {Herminghaus}},\ }\bibfield  {title} {\bibinfo {title} {Electrostatic
  stabilization of fluid microstructures},\ }\href@noop {} {\bibfield
  {journal} {\bibinfo  {journal} {Appl.\ Phys.\ Lett.}\ }\textbf {\bibinfo
  {volume} {81}},\ \bibinfo {pages} {2302} (\bibinfo {year}
  {2002})}\BibitemShut {NoStop}%
\bibitem [{\citenamefont {Liu}\ \emph {et~al.}(2012)\citenamefont {Liu},
  \citenamefont {Wang}, \citenamefont {Chen},\ and\ \citenamefont
  {Robbins}}]{liu2012uncovering}%
  \BibitemOpen
  \bibfield  {author} {\bibinfo {author} {\bibfnamefont {J.}~\bibnamefont
  {Liu}}, \bibinfo {author} {\bibfnamefont {M.}~\bibnamefont {Wang}}, \bibinfo
  {author} {\bibfnamefont {S.}~\bibnamefont {Chen}},\ and\ \bibinfo {author}
  {\bibfnamefont {M.~O.}\ \bibnamefont {Robbins}},\ }\bibfield  {title}
  {\bibinfo {title} {Uncovering molecular mechanisms of electrowetting and
  saturation with simulations},\ }\href@noop {} {\bibfield  {journal} {\bibinfo
   {journal} {Phys.\ Rev.\ Lett.}\ }\textbf {\bibinfo {volume} {108}},\
  \bibinfo {pages} {216101} (\bibinfo {year} {2012})}\BibitemShut {NoStop}%
\bibitem [{\citenamefont {Klarman}\ \emph {et~al.}(2011)\citenamefont
  {Klarman}, \citenamefont {Andelman},\ and\ \citenamefont
  {Urbakh}}]{klarman2011model}%
  \BibitemOpen
  \bibfield  {author} {\bibinfo {author} {\bibfnamefont {D.}~\bibnamefont
  {Klarman}}, \bibinfo {author} {\bibfnamefont {D.}~\bibnamefont {Andelman}},\
  and\ \bibinfo {author} {\bibfnamefont {M.}~\bibnamefont {Urbakh}},\
  }\bibfield  {title} {\bibinfo {title} {A model of electrowetting, reversed
  electrowetting, and contact angle saturation},\ }\href@noop {} {\bibfield
  {journal} {\bibinfo  {journal} {Langmuir}\ }\textbf {\bibinfo {volume}
  {27}},\ \bibinfo {pages} {6031} (\bibinfo {year} {2011})}\BibitemShut
  {NoStop}%
\bibitem [{\citenamefont {Chevalliot}\ \emph {et~al.}(2012)\citenamefont
  {Chevalliot}, \citenamefont {Kuiper},\ and\ \citenamefont
  {Heikenfeld}}]{chevalliot2012experimental}%
  \BibitemOpen
  \bibfield  {author} {\bibinfo {author} {\bibfnamefont {S.}~\bibnamefont
  {Chevalliot}}, \bibinfo {author} {\bibfnamefont {S.}~\bibnamefont {Kuiper}},\
  and\ \bibinfo {author} {\bibfnamefont {J.}~\bibnamefont {Heikenfeld}},\
  }\bibfield  {title} {\bibinfo {title} {Experimental validation of the
  invariance of electrowetting contact angle saturation},\ }\href@noop {}
  {\bibfield  {journal} {\bibinfo  {journal} {J. Adhe. Sci. Technol.}\ }\textbf
  {\bibinfo {volume} {26}},\ \bibinfo {pages} {1909} (\bibinfo {year}
  {2012})}\BibitemShut {NoStop}%
\bibitem [{\citenamefont {Papathanasiou}\ \emph {et~al.}(2008)\citenamefont
  {Papathanasiou}, \citenamefont {Papaioannou},\ and\ \citenamefont
  {Boudouvis}}]{papathanasiou2008illuminating}%
  \BibitemOpen
  \bibfield  {author} {\bibinfo {author} {\bibfnamefont {A.~G.}\ \bibnamefont
  {Papathanasiou}}, \bibinfo {author} {\bibfnamefont {A.~T.}\ \bibnamefont
  {Papaioannou}},\ and\ \bibinfo {author} {\bibfnamefont {A.~G.}\ \bibnamefont
  {Boudouvis}},\ }\bibfield  {title} {\bibinfo {title} {Illuminating the
  connection between contact angle saturation and dielectric breakdown in
  electrowetting through leakage current measurements},\ }\href@noop {}
  {\bibfield  {journal} {\bibinfo  {journal} {J.\ Appl.\ Phys.}\ }\textbf
  {\bibinfo {volume} {103}},\ \bibinfo {pages} {034901} (\bibinfo {year}
  {2008})}\BibitemShut {NoStop}%
\bibitem [{\citenamefont {Peykov}\ \emph {et~al.}(2000)\citenamefont {Peykov},
  \citenamefont {Quinn},\ and\ \citenamefont
  {Ralston}}]{peykov2000electrowetting}%
  \BibitemOpen
  \bibfield  {author} {\bibinfo {author} {\bibfnamefont {V.}~\bibnamefont
  {Peykov}}, \bibinfo {author} {\bibfnamefont {A.}~\bibnamefont {Quinn}},\ and\
  \bibinfo {author} {\bibfnamefont {J.}~\bibnamefont {Ralston}},\ }\bibfield
  {title} {\bibinfo {title} {Electrowetting: a model for contact-angle
  saturation},\ }\href@noop {} {\bibfield  {journal} {\bibinfo  {journal}
  {Colloid Polym.\ Sci.}\ }\textbf {\bibinfo {volume} {278}},\ \bibinfo {pages}
  {789} (\bibinfo {year} {2000})}\BibitemShut {NoStop}%
\bibitem [{\citenamefont {Lubarda}\ and\ \citenamefont
  {Talke}(2011)}]{Lubarda2011}%
  \BibitemOpen
  \bibfield  {author} {\bibinfo {author} {\bibfnamefont {V.~A.}\ \bibnamefont
  {Lubarda}}\ and\ \bibinfo {author} {\bibfnamefont {K.~A.}\ \bibnamefont
  {Talke}},\ }\bibfield  {title} {\bibinfo {title} {Analysis of the equilibrium
  droplet shape based on an ellipsoidal droplet model},\ }\href@noop {}
  {\bibfield  {journal} {\bibinfo  {journal} {Langmuir}\ }\textbf {\bibinfo
  {volume} {27}},\ \bibinfo {pages} {10705} (\bibinfo {year}
  {2011})}\BibitemShut {NoStop}%
\bibitem [{\citenamefont {Roura}\ and\ \citenamefont {Fort}(2004)}]{Roura2004}%
  \BibitemOpen
  \bibfield  {author} {\bibinfo {author} {\bibfnamefont {P.}~\bibnamefont
  {Roura}}\ and\ \bibinfo {author} {\bibfnamefont {J.}~\bibnamefont {Fort}},\
  }\bibfield  {title} {\bibinfo {title} {Local thermodynamic derivation of
  young's equation},\ }\href@noop {} {\bibfield  {journal} {\bibinfo  {journal}
  {J.\ Colloid Interface Sci.}\ }\textbf {\bibinfo {volume} {272}},\ \bibinfo
  {pages} {420} (\bibinfo {year} {2004})}\BibitemShut {NoStop}%
\bibitem [{\citenamefont {Vo}\ \emph {et~al.}(2020)\citenamefont {Vo},
  \citenamefont {Fujita}, \citenamefont {Tagawa},\ and\ \citenamefont
  {Tran}}]{Vo2020anisotropic}%
  \BibitemOpen
  \bibfield  {author} {\bibinfo {author} {\bibfnamefont {Q.}~\bibnamefont
  {Vo}}, \bibinfo {author} {\bibfnamefont {Y.}~\bibnamefont {Fujita}}, \bibinfo
  {author} {\bibfnamefont {Y.}~\bibnamefont {Tagawa}},\ and\ \bibinfo {author}
  {\bibfnamefont {T.}~\bibnamefont {Tran}},\ }\bibfield  {title} {\bibinfo
  {title} {Anisotropic behaviours of droplets impacting on dielectrowetting
  substrates},\ }\href@noop {} {\bibfield  {journal} {\bibinfo  {journal} {Soft
  Matter}\ ,\ \bibinfo {pages} {2621}} (\bibinfo {year} {2020})}\BibitemShut
  {NoStop}%
\end{thebibliography}%

\end{document}